# Charge density waves in YBa$_2$Cu$_3$O$_{6.67}$ probed by resonant x-ray scattering under uniaxial compression

H.-H. Kim[1], E. Lefrançois[1], K. Kummer[2], R. Fumagalli[3], N. B. Brookes[2], D. Betto[1,2], S. Nakata[1], M. Tortora[1], J. Porras[1], T. Loew[1], M. E. Barber[4], L. Braicovich[2,3], A.P. Mackenzie[4,6], C. W. Hicks[4], B. Keimer[1], M. Minola[1]*, and M. Le Tacon[7]†

[1]*Max Planck Institute for Solid State Research, Heisenbergstraße 1, D-70569 Stuttgart, Germany*
[2] *ESRF, The European Synchrotron, 71 Avenue des Martyrs, 38043 Grenoble, France.*
[3]*Dipartimento di Fisica, Politecnico di Milano, I-20133 Milano, Italy.*
[4]*Max Planck Institute for Chemical Physics of Solids, Nöthnitzer Straße 40, 01187 Dresden, Germany*
[5]*CNR-SPIN, Dipartimento di Fisica, Politecnico di Milano, I-20133 Milano, Italy.*
[6]*Scottish Universities Physics Alliance, School of Physics and Astronomy, University of St Andrews, St Andrews KY16 9SS, UK*
[7]*Institute for Quantum Materials and Technologies, Karlsruhe Institute of Technology, D-76344, Eggenstein-Leopoldshafen, Germany*

We report a comprehensive Cu $L_3$-edge resonant x-ray scattering (RXS) study of two- and three-dimensional (2D and 3D) incommensurate charge correlations in single crystals of the underdoped high-temperature superconductor YBa$_2$Cu$_3$O$_{6.67}$ under uniaxial compression up to 1% along the two inequivalent Cu-O-Cu bond directions (*a* and *b*) in the CuO$_2$ planes. We confirm the strong in-plane anisotropy of the 2D charge correlations and observe their symmetric response to pressure: pressure along *a* enhances correlations along *b*, and vice versa. Our results imply that the underlying order parameter is uniaxial. In contrast, 3D long-range charge order is only observed along *b* in response to compression along *a*. Spectroscopic RXS measurements show that the 3D charge order resides exclusively in the CuO$_2$ planes and may thus be generic to the cuprates. We discuss implications of these results for models of electronic nematicity and for the interplay between charge order and superconductivity.

Research on quantum materials aims to develop a quantitative understanding of the macroscopic electronic properties based on atomic-scale correlations between electrons [1]. A powerful experimental strategy is to monitor the response of the macroscopic and microscopic properties of interest as a function of an external control parameter [2]. In research on copper oxide high- temperature superconductors (HTS), recent experiments have indeed uncovered striking anti-parallels between the response of superconductivity measured by macroscopic transport experiments and nm-scale electronic charge correlations measured by x-ray scattering [3-9] to changes in temperature and magnetic field. Increasing the field, for instance, destabilizes superconductivity and enhances the amplitude and correlation length of the charge fluctuations [4,10]. Subsequent work has accumulated evidence for a scenario in which an incommensurate two-dimensional (2D) charge density wave (CDW) reconstructs the Fermi surface [11-14]. This association remains tentative [15], however, because basic questions about the geometry of the Fermi surface in high magnetic fields [16] and the nature of the CDW remain open. Unresolved issues include the directionality (*i.e.*, uniaxial versus biaxial) of the CDW [17,18] and its relationship to electronic "nematicity" (*i.e.*, spontaneously broken lattice-rotational symmetry in the absence of translational symmetry breaking) seen in other experiments [19-23]. These issues also bear directly on the more general question to what extent incommensurate charge correlations are responsible for the anomalous transport and thermodynamic properties in the normal state of cuprates.

To address these questions, we have performed resonant inelastic x-ray scattering (RIXS) experiments in the underdoped HTS $YBa_2Cu_3O_{6+x}$ ($YBCO_{6+x}$) under uniaxial pressure along the Cu-O-Cu bond directions, which couples directly to nematic and uniaxial CDW order parameters. $YBCO_{6+x}$ has served as a model compound for such studies because of its exceptionally large CDW correlation length [3-5,24] and electronic mean free path in quantum transport experiments [12], which greatly facilitate theoretical modelling. The relatively benign influence of doping-induced disorder in $YBCO_{6+x}$ originates in the chain-like order of oxygen dopants in the CuO charge reservoir [25], which orthorhombically distorts its crystal structure (*a* and *b*, perpendicular and parallel to the CuO chains, respectively). Prior non-resonant x-ray scattering (NRXS) experiments on $YBCO_{6+x}$ crystals have probed the response of *b*-axis charge correlations to uniaxial pressure applied along the *a*-axis [26]. At ~1% compression, a new set of Bragg reflections was observed, which could be attributed to a long-range ordered 3D-CDW strongly competing with superconductivity. A related 3D-CDW has also been induced by large magnetic fields [27-29], and its interplay with the shorter range 2D-CDW remains unresolved. A clear benefit of stress experiments over those performed under high fields is the absence of a vortex lattice and associated disorder, as well as the avoidance of the technical limitations associated with implementing x-ray scattering experiments in high fields. Nevertheless, the response of the system in the direction of the applied pressure or the response to pressure along *b* could not be investigated with NRXS due to interference from diffraction signatures of the chain order. These experiments were thus unable to conclusively discriminate between uniaxial and biaxial charge modulations, and to assess differences in the response to pressure along and perpendicular to the modulation direction. We have overcome this limitation by taking advantage of Cu L-edge RIXS, which greatly enhances the sensitivity to valence electrons from the electronically active $CuO_2$ planes with respect to the core electrons that dominate the NRXS signal [30]. In addition, the spectroscopic capability of RIXS allowed us to investigate whether the electrons residing on the CuO chains participate in the 3D-CDW, which was recently reported in experiments on $YBCO_{6+x}$ thin films in the absence of both stress and magnetic field [31].

The RIXS experiments were performed at beamline ID32 of the European Synchrotron Radiation Facility at the Cu $L_3$ absorption edge (932 eV) and with a 60 meV energy resolution [32]. Only resolution-limited elastic scattering representative of the CDW state will be discussed here (full spectra can be found in [33]). Figure 1a shows the scattering geometry. The sample stage was modified to host a strain device [33] similar to the one used in previous work [26,38]. All data presented here were obtained using σ-polarization of the incident x-ray beam to maximize the signal from charge correlations [3].

Twin-free single crystals of $YBCO_{6.67}$ (corresponding to the composition with maximal amplitude and correlation length of the 2D-CDW [5]) were cut into a bar shape and subsequently polished to obtain dimensions of ~1.5 x 0.2 x 0.07 $mm^3$, with the longest dimension along the pressure direction and the shortest along the *c*-axis [33]. In this work, 12 samples were investigated, ensuring high reproducibility of the results. The small size of the incident x-ray beam (4 x 40 $\mu m^2$) compared to the free sample area (900 x 200 $\mu m^2$) ensured homogeneous strain within the illuminated sample volume. During the measurement, the strain level was controlled *in-situ* by monitoring the capacitance of a built-in parallel-plate capacitor. Moreover, we checked the position of the (0, 0, 2) Bragg reflection at each strain level, which shows an increase of the *c*-lattice constant due to the Poisson-ratio expansion under in-plane compression [26,33].

We first focus on the 2D-CDW, which is present in $YBCO_{6.67}$ at temperatures below ~150 K [3,5]. Figure 1b,c shows scans around the incommensurate CDW wavevectors $\mathbf{Q}^a_{//} = (0.305, 0)$ and $\mathbf{Q}^b_{//} = (0, 0.315)$ in the two-dimensional Brillouin zone in the absence of pressure (The components *H*, *K*, *L* of **Q** are given in reciprocal-lattice units - *L* was set close to 1.5, where the 2D-CDW intensity is maximal [4,26]). To accurately determine the lineshape of the CDW satellites, we took advantage of the four-circle sample goniometer on ID32 that allows scans along the (*H*, 0, 0) and (0, *K*, 0) directions in reciprocal space, irrespective of the direction of the applied pressure, and allowed us to overcome limitations of previous studies [3,5,17]. From the widths of the line scans, we can directly extract the longitudinal ($\xi^a_{//}$, $\xi^b_{//}$) and transverse ($\xi_\perp^a$, $\xi_\perp^b$) correlation lengths of the real-space 2D-CDW domains in the *ab*-plane. In the absence of stress, we confirm the anisotropic shape of the CDW satellites [17]. This anisotropy does not seem to originate from the orthorhombic distortion of underdoped YBCO (see [17] and discussion below), and might instead arise from the oxygen ordering and related defects in the chain layer, potentially limiting the CDW correlation length. This is corroborated by the fact that the oxygen-VIII order in YBCO$_{6.67}$ has a longer correlation length along b than a ($\xi^a \sim 12$ Å, $\xi^b \sim 51$ Å), comparable with that of the CDW. Following Ref. [17] and for simplicity, we represent this anisotropy in the form of ellipses in Fig. 1b, c (border thicknesses reflect weak sample-to-sample variations). We cannot rule out that the actual shape deviates from ellipses, but this would not affect our conclusions. As argued in [17], the observation of four different correlation lengths suggests a real-space picture where unidirectional CDWs form domains with charge modulation directions along either *a*- or *b*-axes, *i.e.* stripes of uniform charge density along either *b* or *a* (Figure 1d). We refer to these domains as *a*-CDW and *b*-CDW, respectively. However, an alternative analysis of this pattern in terms of a "checkerboard" CDW was also proposed [15].

To distinguish between these scenarios, we now turn to the effect of uniaxial pressure. The panels a-d in Figure 2 highlight a striking symmetry in the pressure dependence of the CDW: pressure along *a* enhances the intensity and reduces the width of the *b*-CDW reflections but leaves those of the *a*-CDW nearly unaffected, and vice versa. This behavior shows that the *a*-CDW and the *b*-CDW are distinct order parameter components, that evolve independently under uniaxial pressure. Our data firmly rule out that the 2D-CDW can be described by a single checkerboard order parameter. This conclusion is based on the differential response to uniaxial pressure and is independent of the shape of the uniaxial domains, which evolve differently along the two perpendicular axes (Figs. 2e,f and Fig. 2g,h in momentum and real space, respectively). The faster growth of the *b*-CDW domain size in response to *a*-axis stress agrees qualitatively with the behavior of the superconducting $T_c$, which is rapidly suppressed with *a*-axis stress but much less susceptible to stress along *b* [33,39,40]. This observation underscores the competition between superconductivity and CDW. We also note that the symmetry of the pressure response rules out an essential role of the built-in difference between the lengths of *a*- and *b*-axis, which is enhanced (reduced) by *a*-axis (*b*-axis) compression, respectively.

Prior work including particularly the observation of low-energy phonon anomalies at $\mathbf{Q}^b_{/\!/}$ [41] has shown that the diffraction signal under ambient conditions can be understood as a "central peak" resulting from pinning of low-energy collective charge density fluctuations by defects. This insight suggests that the pronounced overall increase of the elastic intensity under modest uniaxial pressure is fed from the condensation of such fluctuations. Within experimental accuracy, a 1% compression along *a*-axis (*b*-axis) yields a doubling of the elastic intensity of the *b*-CDW (*a*-CDW), respectively (Figs. 2b,c). This symmetry then implies that these fluctuations are of comparable strength along *a* and *b*. It is interesting to consider the implications of this observation for the hypothesis of a nematic order parameter in YBCO$_{6+x}$ and other cuprates, which was motivated in part by emergent transport anisotropies in the pseudogap regime [19,21]. The *a*/*b* symmetric fluctuations do not *per se* rule out a spontaneously broken lattice-rotation symmetry, as this pattern can also be generated by an equal population of nematic domains. This would then imply that the orthorhombic distortion of the YBCO$_{6+x}$ crystal structure does not align the nematic order parameter in the charge sector – in stark contrast to the spontaneous alignment of the propagation vector of incommensurate spin fluctuations below ~150 K, which was discussed as evidence of nematicity [20]. The nematic susceptibilities in the charge and spin sectors thus appear to be decoupled, and at least on the qualitative level, the spontaneous transport anisotropies [19,21] show greater parallels with spin rather than charge correlations.

Our comprehensive study of uniaxial pressure effects also has interesting implications for the 3D-CDW recently discovered in NRXS experiments on YBCO$_{6.67}$ with pressure along *a* [26]. Figure 3a, c shows the pressure dependence of the intensity at 3D-CDW Bragg peak positions at $L \sim 1$. Our RIXS data confirm a clear enhancement of the signal at $\mathbf{Q}^b_{//}$ as the *a*-axis is compressed by ~1%, (Fig. 3a). We note that the observed peak is not as intense as that seen in NRXS. In addition to possible weaker structure factor of the 3D-CDW at the momentum-space positions accessible to Cu-$L_3$ RIXS [29], the peak is strongly distorted by self-absorption effects and does not reflect the long-range nature of the order [33]. The temperature dependence of the Bragg intensity, which is sharply peaked at the superconducting $T_c$ (Fig. 3e) also agrees with the NRXS data [26].

Our RIXS experiments allowed us to extend our knowledge of the 3D-CDW by studying the effect of *b*-axis compression (Figs. 3c). The resulting data show that *b*-axis pressure does not generate any discernible enhancement of the $L \sim 1$ signal at $\mathbf{Q}^a_{//}$. The apparent lack of a 3D-CDW propagating along *a*, at least at the currently accessible pressure levels, departs from the symmetry exhibited by the response of the 2D-CDW (Figs. 2a-d). Having ruled out the intrinsic orthorhombic strain as possible origins of this difference, we consider two alternative mechanisms. First, the 3D-CDW might arise when growth of the 2D-CDW correlation length effectively enhances the interaction among 2D-CDWs in adjacent bilayers [42,43], in a manner similar to the mechanism leading to 3D antiferromagnetic order in undoped cuprates [44]. However, we have seen that the *a*-CDW domains are generally larger than the *b*-CDW ones, and that the 2D-CDW domains actually grow faster with pressure along *a* than along *b*, in contradiction to this hypothesis. Further evidence against this scenario comes from the lack of an intensity reduction of the 2D-CDW signal at the critical pressure for 3D-CDW formation, which was also noted for the magnetic-field-induced 3D-CDW [28,29,45].

Another factor that might favor a 3D-CDW along *b* is the influence of the CuO chains valence electrons. Recent RIXS experiments on underdoped YBCO$_{6+x}$ thin films [31] have revealed Bragg reflections with resonant enhancement at the characteristic energy of the chains Cu sites, suggesting direct participation of these electrons in the 3D-CDW. In contrast, our 3D-CDW feature strongly peaks at the in-plane Cu resonance, just like the 2D-CDW, which indicates that the 3D-CDW order in films and in strained single crystals is qualitatively different. The different temperature dependence of the 3D-CDW in both systems underscores this conclusion. The reflection intensity in the strained crystal (Fig. 3e) indeed sharply peaks at $T_c$ and vanishes at ~75 K, whereas the one in the films is hardly affected by the onset of superconductivity and persists up to room temperature without noticeable anomalies.

In bulk YBCO$_{6.67}$, valence electrons on the CuO chains do not participate directly in the 3D-CDW. Nevertheless, as they do affect the shape of the Fermi surface via hybridization between energy bands derived from the CuO$_2$ planes and the CuO chains, they may thus indirectly influence the propensity of the electron system in the CuO$_2$ planes to CDW formation. The importance of electron-phonon interactions (EPI) for the formation of the CDW in the cuprates has been highlighted [9,26,41,46-49], and research on classical CDW compounds has shown that its momentum dependence is also a critical parameter [50-52]. We argue here that the pronounced a/b anisotropies of EPI for high-energy phonons reported in YBCO [53,54] reflect the influence of the CuO chains on the EPI and play an important role in destabilizing the 3D *a*-CDW relative to the *b*-CDW. In contrast, in effectively tetragonal YBCO$_{6+x}$ films, the absence of oxygen order may render these destabilizing effects less effective and hence contribute to the higher stability of the 3D-CDW [31].

We end our discussion by laying out a concrete scenario for the interplay between 2D-CDW, 3D-CDW, and superconductivity that is consistent with all data collected under uniaxial stress. For T ≲ 150 K, the 2D-CDW is nucleated by randomly placed defects and remains short-range-ordered under any combination of temperature [3,6-8,55], pressure [26], and magnetic field [4,29,45]. The natural defect site is the off-stoichiometric CuO chain layer [25], which is located half-way between adjacent CuO$_2$ bilayers. Defects in the chain layers will thus force in-phase alignment of CDWs in directly adjacent CuO$_2$ layers that belong to bilayer units in different crystallographic unit cells. The effective unit-cell doubling of the short-range-ordered state (indicated by maxima of the 2D-CDW reflections for half-integer *L*) can be attributed to Coulomb repulsion, which favors anti-phase correlation between uniaxial CDWs within a single bilayer unit [43]. The 2D-CDW is weakened by the onset of superconductivity, but persists locally because of pinning to defects.

The 3D-CDW, on the other hand, preserves the periodicity of the crystal lattice along *L*, likely due to electrostatically driven phase alternation of uniaxial CDWs both within a bilayer and between neighboring bilayers [43]. When sufficiently large *a*-axis pressure is applied, the 3D-CDW is thermodynamically stable below ~75 K even in the absence of defects, and thus forms a long-range ordered state in regions that are not already covered by 2D-CDW patches. However, superconductivity fully supplants the 3D-CDW upon cooling below ~55 K, in what can be regarded as a manifestation of thermodynamic competition between two bulk phases with genuine long-range order.

In conclusion, the response of the RIXS elastic signal to uniaxial compression unambiguously demonstrates the uniaxial nature of 2D and 3D charge order in $YBCO_{6+x}$. RIXS data at high pressure further show that the 3D-CDW resides entirely in the $CuO_2$ planes and may thus be generic to the cuprates. Investigations to check whether this order can also be induced in other families of cuprates, encompassing Hg- [8,46] or Bi-based [6,7] ones are highly desired. The novel methodology we have introduced – RIXS under uniaxial pressure – opens up avenues for investigation in many other materials.

**Acknowledgements**

The experimental data were collected at the beam line ID32 of the European Synchrotron (ESRF) in Grenoble (France) using the ERIXS spectrometer through the beamtime allocated within the Long Term Project HC-2602 *Resonant Soft X-ray Spectroscopy under Uniaxial Strain.* We are grateful to G. Ghiringhelli, C. Meingast, Y. Peng, J. Schmalian, S. M. Souliou and R. Willa for fruitful discussions. R.F acknowledges support by Fondazione CARIPLO and Regione Lombardia under project "ERC-P-ReXS" No. 2016-0790. MLT and CWH This work was funded by the Deutsche Forschungsgemeinschaft (DFG, German Research Foundation) - TRR288 - 422213477 (project B03).

*m.minola@mpi.fkf.de, †matthieu.letacon@kit.edu

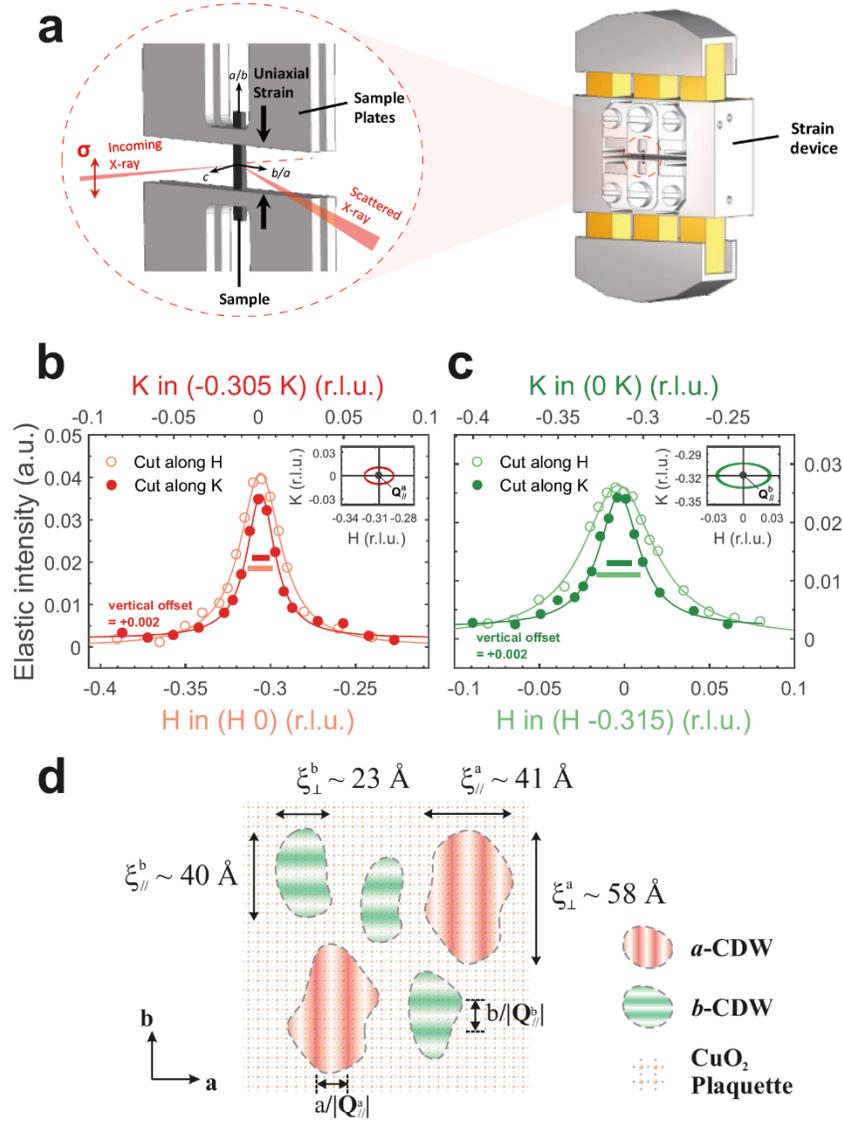

**Figure 1** (a) Scattering geometry with the strain device. (b,c) Strain-free quasi-elastic intensity scans of the *a*-CDW *(b*-CDW) propagating along *a*- (*b*-) axis across $\mathbf{Q}^a_{//}$ = (0.305 0) ($\mathbf{Q}^b_{//}$=(0 0.315)), respectively. Light (dark) colors represent longitudinal (transverse) scans (T= 55K). Inset of (b,c): Momentum-space shape of the CDW reflections. The semi-major (-minor) axes represent the half-widths-at-half-maximum of the reflections along *H* (*K*). (d) Real-space cartoon of CDW domains with different modulation directions in the CuO$_2$ plane. $\xi^a_{//}$ and $\xi^b_{//}$ ($\xi^a_\perp$ and $\xi^b_\perp$) are longitudinal (transverse) correlation lengths of the *a*- and *b*-CDW, respectively.

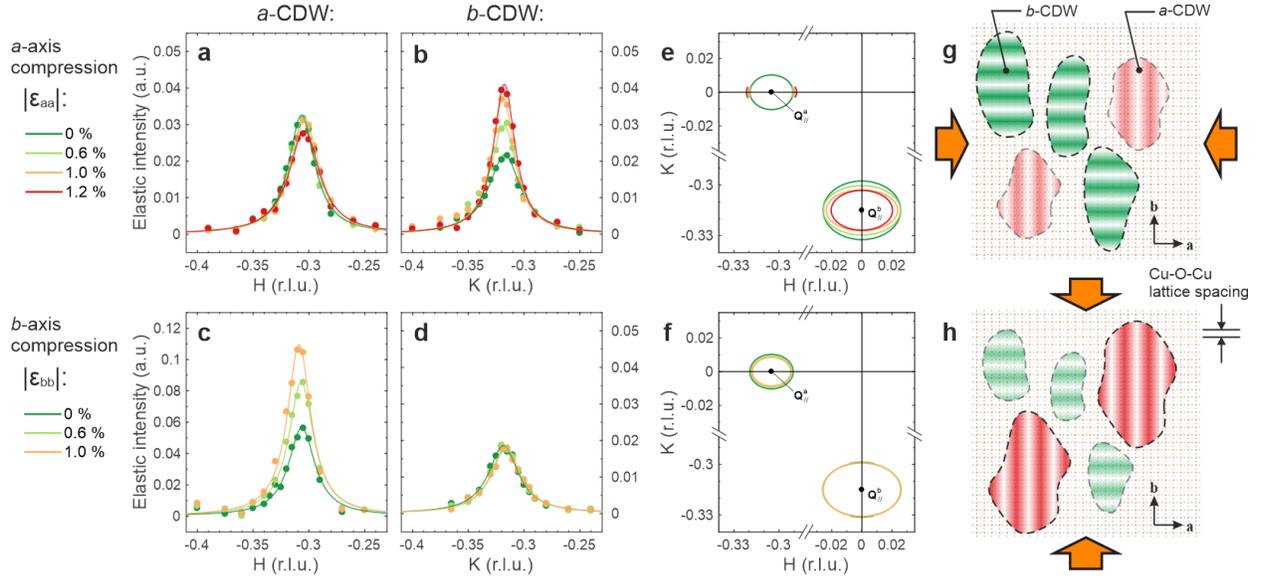

**Figure 2** (a-d) Pressure dependence of the 2D-CDW at $\mathbf{Q}^a_{//}$ ($\mathbf{Q}^b_{//}$) under (a,b) *a*-axis and (c,d) *b*-axis compression (T = 55 K). The lines are Lorentzian fits. (e,f) Evolution of the reflection widths under (e) *a*-axis and (f) *b*-axis compression. Semi-major (-minor) axes of the half-widths-at-half-maximum of the peaks along *H* (*K*). (g,h) Real-space pictures of CDW domains in the CuO$_2$ plane under (g) *a*-axis and (h) *b*-axis compression.

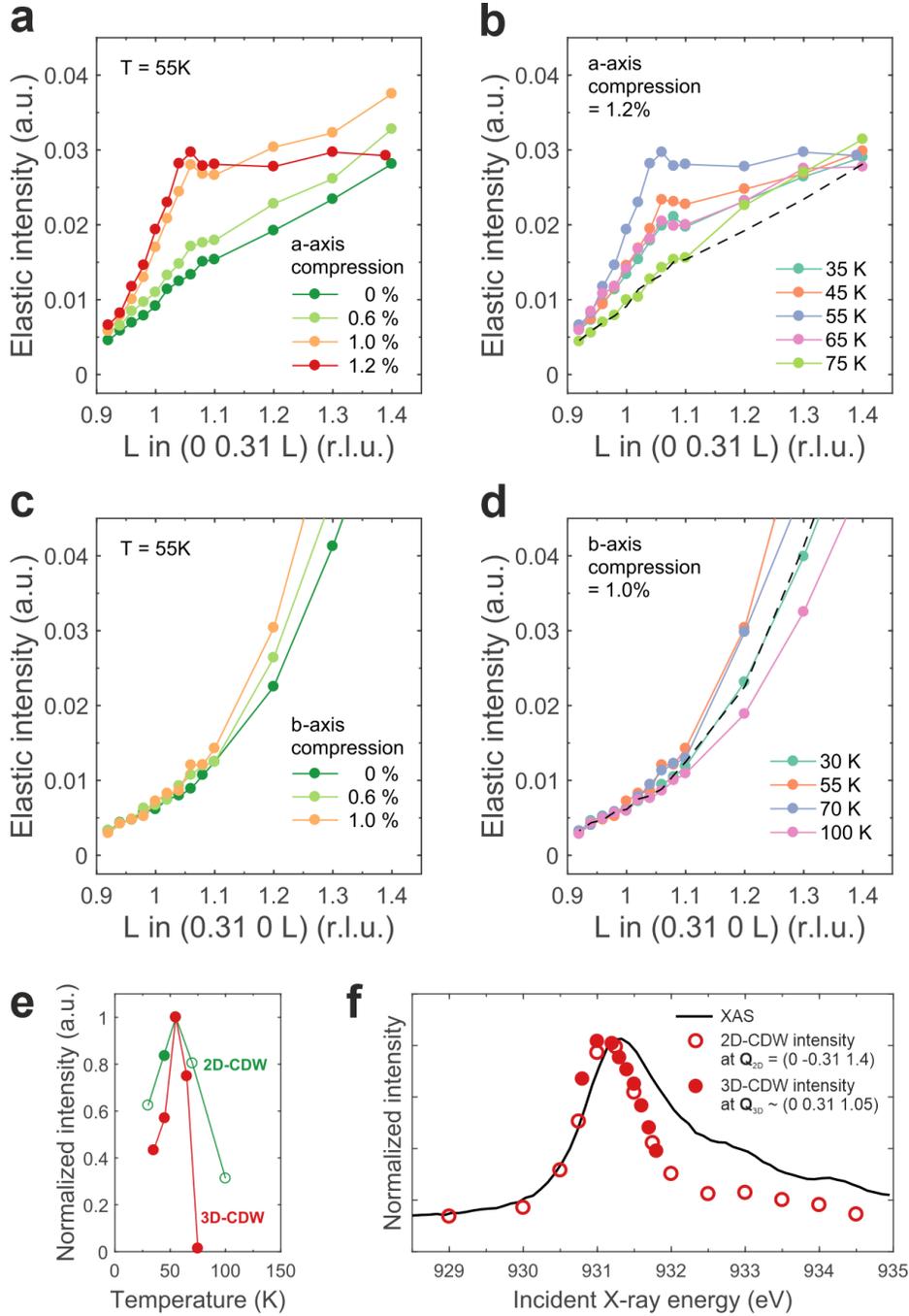

**Figure 3** (a-d) 3D-CDW intensity at $\mathbf{Q}^a_{//}$ ($\mathbf{Q}^b_{//}$) and $L\sim1$ under (a) *a*-axis compression (T=55K), (b) 1.2% *a*-axis compression, (c) *b*-axis compression at T=55K, (d) 1.0% *b*-axis compression. Lines are guides to the eye. Black dashed lines in (b,d) show strain-free reference (T=55K). Background levels of (c,d) for L > 1.1 are larger than that of (a,b) since it also involves substantial background signals from the oxygen chain peak. (e) Temperature dependence (normalized to T=55K) of the integrated intensity of 2D-CDW and 3D-CDW (Red circles, measured at 1.2% *a*-axis compression) reflections. Empty (filled) green circles are intensities of reflections at $\mathbf{Q}^a_{//}$ ($\mathbf{Q}^b_{//}$) measured at 1.2% (1.0%) *a*- (*b*-) axis compression. (f) Comparison between the x-ray absorption spectrum (black line) and the photon energy dependence of the peak intensities of 2D- and 3D-CDW reflections (T = 55 K).

13